\documentstyle[prd,aps,preprint,tighten,epsfig]{revtex}

\begin{document}

\draft

\title{The Gross--Llewellyn Smith Sum Rule
in the Analytic Approach to Perturbative QCD}

\author{K.A. Milton\thanks{E-mail: milton@mail.nhn.ou.edu}}

\address{Department of Physics and Astronomy,
University of Oklahoma, Norman, OK 73019 USA}

\author{I.L. Solovtsov\thanks{E-mail: solovtso@thsun1.jinr.ru} and
O.P. Solovtsova\thanks{E-mail: olsol@thsun1.jinr.ru}}
\address{Bogoliubov Laboratory of Theoretical Physics,
 Joint Institute for Nuclear Research, 141980 Dubna, Moscow Region, Russia }
\preprint{OKHEP--98--07}
\date{\today}
\maketitle

\begin{abstract}
We apply analytic perturbation theory to the Gross--Llewellyn Smith sum rule.
We study the $Q^2$ evolution and the renormalization scheme dependence of
the analytic three-loop QCD correction to this sum rule,
and demonstrate that the results are practically renormalization scheme
independent and lead to rather different $Q^2$ evolution
than the standard perturbative correction possesses.
\end{abstract}


\newpage

\section{Introduction}

At present two deep inelastic scattering (DIS) sum rules,
the Gross--Llewellyn Smith (GLS)~\cite{GLS} and the polarized Bjorken sum
rule~\cite{Bj}, give the possibility of extracting the
value of the strong coupling constant $\alpha_S$ from experimental data,
in particular, at low momentum transfer, $Q$, down to $1\,{\mbox{\rm GeV}}$
(see, e.g., Refs.~\cite{Burrows,Albrow}).
Comparison of these values of $\alpha_S$ with other accurate
$\alpha_S$ values, such as those obtained from the decay widths of the
$\tau$-lepton and the $Z$-boson into hadrons, is an important test
of the consistency of QCD. Clearly what are required are reliable theoretical
relations between physically observable quantities and the strong coupling
constant $\alpha_S$. At low $Q^2$ scales,
there are significant theoretical uncertainties,
which come, firstly, from a truncation of the series obtained from
perturbation theory (PT), which leads to a significant renormalization
scheme~(RS) ambiguity, and, secondly, from poorly understood nonperturbative
effects (see, e.g., Refs.~\cite{Shifman,Fischer} for a review).
In this paper we apply the method proposed in Refs.~\cite{DVS,DVS2}
(see also  Refs.~\cite{KS1,DV}), the so-called analytic perturbation
theory (APT), to study the GLS sum rule, continuing our investigation of
the APT approach initiated in Refs.~\cite{MSS1,KO1}.
This method takes into account the fundamental principle of causality,
which in the simplest cases is reflected in the form of $Q^2$-analyticity
of the K\"all$\acute{{\rm e}}$n--Lehmann type.
The standard renormalization group resummation violates this required
analytic structure, and unphysical singularities such as a ghost pole appear.
The APT approach maintains the causal properties and removes all unphysical
singularities by incorporating nonperturbative terms (for the number
of flavors occurring in nature) to secure the required analytic properties.
These nonperturbative terms can be presented as power corrections and their
appearance is not inconsistent with the operator product
expansion~\cite{Grunberg97}.

It is familiar that the QCD correction to the GLS sum rule is
expressed as an expansion in powers of $\alpha_S$
that allows one, in principle, easily to obtain a value for the
running coupling constant. Before going into detailed considerations,
let us demonstrate the difference between the standard PT and APT
running coupling constants in the two-loop level. The two-loop analytic
running coupling constant can be written in the form of a sum
of the standard perturbative part and additional terms which compensate for
the contributions of unphysical singularities, the ghost pole and a cut
arising from the ``log-of-log" dependence.
The difference may be transparently shown by using the approximate
formulae for the two-loop analytic running coupling given
in Ref.~\cite{DVS2}
\begin{equation}
\label{APT_app}
{\alpha}_{\rm APT}(Q^2,\Lambda,n_f)\,\simeq \,
{\alpha}_{\rm PT}(Q^2,\Lambda,n_f)\,+\,
\frac{2\pi}{\beta_0(n_f)}\left[
\frac{1}{1-Q^2/\Lambda^2}-2 C_1(n_f)\frac{\Lambda^2}{Q^2} \right] \, ,
\end{equation}
where $\beta_0=11-2n_f/3$  is the one-loop coefficient of the
$\beta$-function corresponding to $n_f$ active quarks,
and, for four active flavors, $C_1(n_f=4)=0.0396 \,.$
The PT running coupling  constant is obtained
by integration of the renormalization group
equation with  the two-loop $\beta$-function.

\begin{figure}[thp]
\centerline{ \epsfig{file=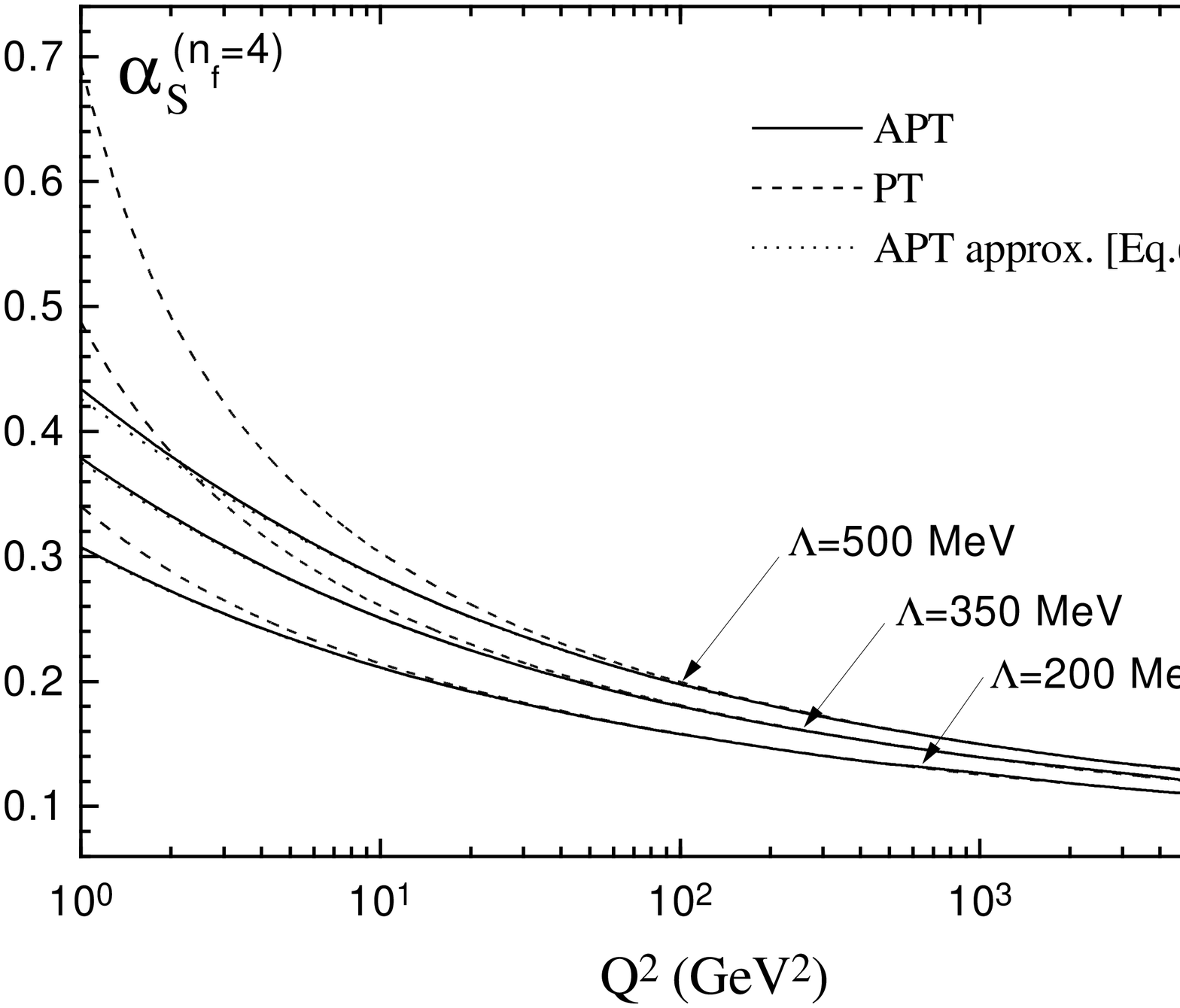,width=10.0cm}}
\caption{Comparison of $Q^2$ evolution of the two-loop
running coupling constant
in the APT and the standard PT approaches.
 }
\label{gls_fig1}
\end{figure}

The difference between the APT and PT functions
is illustrated in Fig.~\ref{gls_fig1} over a wide range of $Q^2$,
$1 \leq Q^2 \leq 10^4\, {\mbox {\rm GeV}}^2$.
The curves in the figure correspond to various values of the QCD
scale parameter $\Lambda$ for four active flavors.
Fig.~\ref{gls_fig1} shows that the expression~(\ref{APT_app}),
represented by dotted lines,
approximates the exact two-loop analytic coupling (solid lines)
rather well for $Q>2\Lambda$.
Comparing the $Q^2$ evolution of the QCD running coupling  constant
obtained from APT to that given by PT~(dashed lines), one can see
that the difference between the shapes of
the APT and PT running coupling constants becomes significant
at low $Q^2$-scales, $ Q^2 < 10\, {\mbox {\rm GeV}}^2$.
This fact stimulated applications of the modified perturbation theory
with correct analytic properties, APT,
for various physical processes (see, e.g., Refs.~\cite{MSS1,MSS2}).
Here, we consider the GLS sum rule in the framework of the
APT approach. As  has been demonstrated in Ref.~\cite{W78}
by using the Deser-Gilbert-Sudarshan representation for the virtual forward
Compton amplitude~\cite{DGS} (see also Ref.~\cite{Ashok_suri}),
the moments of the structure functions are analytic
functions in the complex $Q^2$-plane with a cut along the negative real axis.
On the other hand, the conventional renormalization group resummation
does not support these analytic properties and the influence of requiring
these properties to hold in the DIS description has not
been studied. Here, we perform this investigation, by
applying the APT method, which gives the possibility
of combining the renormalization group resummation
with correct analytic properties of the QCD correction to the GLS
sum rule.
In Sec.~II we start by describing the GLS sum rule in the PT and APT approaches
and compare the $Q^2$ evolution of the APT and PT predictions.
In Sec.~III we consider in detail the RS
dependence of the results. Summarizing comments are given in Sec.~IV.

\section{The GLS sum rule within APT}

The GLS sum rule predicts the value of the integral
over all $x$ of the non-singlet $F_3$ structure function
measured in neutrino- and antineutrino-proton scattering
\begin{equation}
\label{GLS}
S_{\rm GLS}(Q^2)=\frac{1}{2}\int_{0}^{1}dx
\left[F_3^{\nu p}(x,Q^{2})+F_3^{{\bar\nu} p}(x,Q^{2})\right] \,.
\end{equation}
In the quark-parton level, which is appropriate for $Q^2 \to \infty$, the
GLS sum rule should equal three. Therefore, for fixed $Q^2$
the integral~(\ref{GLS}) can be conveniently written as
\begin{equation}
\label{GLS1}
S_{\rm GLS}(Q^2) \,=\,3\,
\left[1\,- \, \Delta_{\rm GLS}(Q^2) \right ] \,,
\end{equation}
where the QCD correction, $\Delta_{\rm GLS}$, in principle, contains
perturbative and nonperturbative parts.
To begin, we concentrate on the perturbative contribution
to $\Delta_{\rm GLS}$, considering in turn standard PT and
APT methods and postponing until later a discussion of
the possible influence of higher twist~(HT) effects,
which remain poorly understood. In this connection
it is interesting to note from the result of the fit to the
revised CCFR data~\cite{Seligman97}
presented in Refs.~\cite{Kataev98} that the higher-twist contributions
are small in the region $Q^2 > 1\, {\mbox {\rm GeV}}^2$
and can have a sign-alternating character.

The standard perturbative part of the GLS sum rule correction
is known up to the three-loop level
for massless quarks in the MS-like renormalization schemes
with the number of active quarks $n_f$ fixed,
\begin{equation}
\label{Delta_PT}
\Delta_{\rm GLS}^{\rm PT} =
\frac{{\alpha}_{\rm PT}}{\pi}\,
+ d_1({\rm RS},n_f)\left(
\frac{{\alpha}_{\rm PT}}{\pi}\right)^2 +
d_2({\rm RS},n_f)
\left( \frac{{\alpha}_{\rm PT}}{\pi}
\right)^3.
\end{equation}
The PT running coupling constant ${\alpha}_{\rm PT}(Q^2,{\rm RS},n_f)$
is obtained by integration of
the renormalization group equation with the three-loop $\beta$-function.
The coefficients $d_1$ and $d_2$ are given in Ref.~\cite{LV91}
in the $\overline{\rm MS}$ renormalization scheme.
Therefore, the perturbative QCD correction to the GLS sum rule is
represented in the form of a power series in ${ \alpha}_{\rm PT}\,$
and, at first glance, the value of ${\alpha}_{\rm PT}\,$
can be easily extracted if the value of $\Delta_{\rm GLS}(Q^2)$ is
experimentally known. In the region $Q^2 > 1\, {\mbox {\rm GeV}}^2$,
one believes PT with its renormalization-group improvement
is still valid.  We should note that different regions of the $x$ integration
in Eq.~(\ref{GLS}) at fixed values of $Q^2$, in principle, correspond to
different numbers of active quarks, $n_f$; arguments have been given
in~Ref.~\cite{CK92} to select one or another value of $n_f$. Experimental
measurements of the GLS sum rule are made in a region of $x$ where one
believes that four light flavors are relevant. At present, there is no
regular and consistent method of including threshold effects for the GLS
sum rule. In the following analysis, we will first take $n_f=4$  to
obtain results in the standard $\overline{\rm MS}$ renormalization scheme,
and then consider RS dependence.

The QCD correction with correct analytic properties can be written
in the form of a spectral representation
\begin{equation}
\label{Delta_ro}
\Delta_{\rm GLS}(Q^2) \, =
\,\frac{1}{\pi}\,
\int_0^\infty\,\frac{d\sigma}{\sigma\,+\,Q^2} \,
\varrho(\sigma) \, ,
\end{equation}
where we have introduced the spectral function, which is defined as
the discontinuity of $\Delta_{\rm GLS}(Q^2)$:
$\,\varrho(\sigma)\,= \,
{\rm Disc}
\left \{\Delta_{\rm GLS}(-\sigma-{\rm i}\epsilon)
\right\}/{2 {\rm i}}\, $.
If we calculate now the spectral function $\,\varrho(\sigma)$ perturbatively,
we get an expression for $\Delta_{\rm GLS}(Q^2)$, which has the correct
analytic properties and therefore no unphysical singularities.
Consequently, we write the three-loop APT approximation
to $\Delta_{\rm GLS}(Q^2)$ as follows
\begin{equation}
\label{Delta_APT}
\Delta_{\rm GLS}^{\rm APT}\,=\,
{\delta_{\rm APT}^{(1)}(Q^2)}
\,+\, d_1 \;{\delta_{\rm APT}^{(2)}(Q^2)}
  \,+ \,
d_2\; {\delta_{\rm APT}^{(3)}(Q^2)}   \,,
\end{equation}
where the coefficients $d_1$ and $d_2$ are the same as
in Eq.~(\ref{Delta_PT}) and
the functions ${\delta_{\rm APT}^{(k)}(Q^2)}$ are derived from the
spectral representation and correspond to
the discontinuity of the $k$-th power of the PT running coupling constant
\begin{equation}
\label{del_APT}
{\delta_{\rm APT}^{(k)}(Q^2)} \,=
\,\frac{1}{\pi^{k+1}}\,
\int_0^\infty\,\frac{d\sigma}{\sigma\,+\,Q^2} \,
 {\rm Im} \;
 \left \{ {\alpha}_{\rm PT}^k(-\sigma -{\rm i}\varepsilon)
 \right\}\,  .
\end{equation}
The function $\delta_{\rm APT}^{(1)}(Q^2)$
defines the APT running coupling constant,
${ \alpha}_{\rm APT}(Q^2) \,=\, \pi \delta_{\rm APT}^{(1)}(Q^2)\,$,
which in the one-loop order is given by
\begin{equation}
\label{APT1}
{ \alpha}_{\rm APT}(Q^2)\,=\,
\, \frac{4\pi}{\beta_0}\,
\left[\, \frac{1}{\ln \left(Q^2/\Lambda^2\right)}\,+\,
\frac{1}{1-Q^2/\Lambda^2} \right ] \>.
\end{equation}
As can be seen from Eq.~(\ref{Delta_APT}), the first term of the expansion is
${ \alpha}_{\rm APT}\,/\pi$, but the following terms are not
representable as powers of ${ \alpha}_{\rm APT}\,$ unlike in the PT case.
There are approximate expressions, like Eq.~(\ref{APT_app}), for
higher loop corrections in Eq.~(\ref{Delta_APT}), which have rather simple
forms, which can be derived by using a method of subtracting unphysical
singularities~\cite{ils_new}. For instance, for $\delta_{\rm APT}^{(2)}$,
the approximate formula is
\begin{equation}
\label{d2_approx}
\delta_{\rm approx}^{(2)}(z)=
\left[\frac{{\alpha}_{\rm PT}(z)}{\pi}\right]^2 +
\frac{4}{\beta_0^2}\left[-\frac{1}{(1-z)^2}
+\left( 1+\frac{\beta_0^2}{2\beta_1}\right)\frac{1}{1-z}
+\frac{C_1}{z}\right]\, ,\, z=Q^2/\Lambda^2 \, ,
\end{equation}
where $C_1=0.0273$ for four active flavors and $\beta_1=102-38n_f/3$.

To illustrate the difference between the convergence properties of the
PT expansion (\ref{Delta_PT}) and the APT series (\ref{Delta_APT})
we use the recent result of the CCFR Collaboration
(CCFR'97)~\cite{Spentzouris,Yu1}:
$S_{\rm GLS}=2.47 \pm 0.09$ at $\,Q^2=3.1\,{\mbox{\rm GeV}}^2$,
which is consistent with the result of a previous
CCFR~analysis (CCFR'93)~\cite{CCFR93},
$S_{\rm GLS}=2.50 \pm 0.08$ at $\,Q^2=3\,{\mbox{\rm GeV}}^2$.
The central value of $S_{\rm GLS}$ corresponds to the
value of the QCD correction $\Delta_{\rm GLS}^{{\rm exp}}=0.1767$
and the successive terms of the PT series~(\ref{Delta_PT})
respectively constitute 65.1\%, 24.4\% and 10.5\% of the total. At the same 
time, the corresponding contributions to the APT series~(\ref{Delta_APT}) make
up 75.7\%, 20.7\% and 3.6\% of the total.
The convergence of the APT series seems to be somewhat better behaved
than is that of the PT expansion at such small
$Q\simeq 1.76\,{\mbox{\rm GeV}}\,$.

\begin{figure}[thp]
\centerline{ \epsfig{file=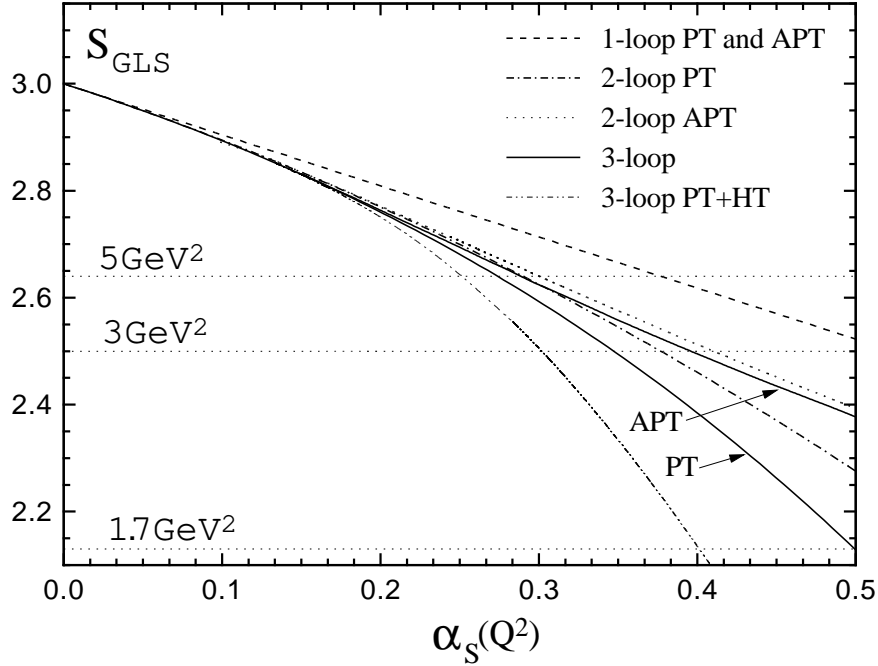,width=10.0cm}}
\caption{$S_{\rm GLS}\,$  with $1$-, $2$-, and $3$-loop
QCD corrections vs. the coupling constant. }
\label{gls_fig2}
\end{figure}

The same may be seen from Fig.~\ref{gls_fig2}, where $S_{\rm GLS}$ is shown
as a function of the QCD running coupling constant $\alpha_{S}$ in the PT
and APT approaches. As outlined above, in the PT case, the function
$S_{\rm GLS}$ is an explicit function of the PT running coupling
constant and in the one-loop approximation is represented by a straight
line in Fig.~\ref{gls_fig2}, as a parabola in the two-loop case,
and as a cubic curve in the three-loop one.
At sufficiently large values of $\alpha_{S} \sim 0.4$, the difference
between the 1-, 2-, and 3-loop PT predictions becomes large.
An inclusion of the higher-twist term with the value recommended by the
Particle Data Group,  $\Delta_{\rm HT}=(0.09\pm0.045)/Q^2$~\cite{PDG96}
(see Refs.~\cite{BBK,Ross} for additional details),
also significantly changes the behavior, as is apparent from the figure.
Note that the coincidence of the one-loop PT and APT curves
in Fig.~\ref{gls_fig2} does not mean that the PT and APT approaches
are physically identical, this is simply a matter of the linear form of
the one-loop approximation; the behavior of the PT and APT running
coupling constants are rather different [see Eq.~(\ref{APT1})].
In the APT case, the contribution of the higher loop corrections is not
so large as in the PT one and the corresponding curves in
Fig.~\ref{gls_fig2} are quite close to the linear function, and, especially,
there is very little difference between the 2-loop and 3-loop results.
The horizontal lines in Fig.~\ref{gls_fig2} correspond
to central values from experimental data at different
low values of $Q^2$:
$S_{\rm GLS}(1.7\,{\mbox{\rm GeV}^2})=2.13
\pm 0.46$~\cite{Barabash},
$S_{\rm GLS}(3\,{\mbox{\rm GeV}^2})=2.50
\pm 0.08$~\cite{CCFR93},
$S_{\rm GLS}(5\,{\mbox{\rm GeV}^2})=2.63
\pm 0.09$~\cite{Spentzouris,Yu1}.
The intersection of one of these lines with
a given theoretical curve gives the value of $\alpha_S$ for that value of
$Q^2$ in that theoretical description.  It will be seen that stability of
the theoretical curve at $Q^2 \sim$~ a~few~GeV$^2$ is required in order
to extract a reliable value of $\alpha_S$ or the QCD scale parameter
$\Lambda$.

Consider the $Q^2$ evolution of the GLS sum rule.
In Ref.~\cite{Harris}, the low-$Q^2$ dependence of GLS sum rule
has been evaluated by combining $xF_3$ measurements of the
CCFR'93 data with data from other DIS experiments.
Preliminary updated analysis with CCFR'97 data was presented
in Refs.~\cite{Spentzouris,Yu1}.
The analysis of the GLS sum rule based on the Jacobi polynomial
expansion has been given in Ref.~\cite{KataevQCD94},
and for new CCFR'97 data has been examined in Ref.~\cite{Sidorov}.
The very recent CCFR/NuTEV result for the GLS integral
is presented in Refs.~\cite{Yu2,ICHEP,Harris_2}.
The three values of the GLS integral at $Q^2=1.26$,
$2.00$, and $3.16$~GeV$^2$ are in good agreement
with the old result~\cite{Harris}, but the value at $Q^2=5$~GeV$^2$ is larger,
which corresponds to a smaller value of the QCD correction to GLS sum rule,
although consistent within errors (see also Fig.~4). Note that the value
$S_{\rm GLS}(Q^2=5.01\,{\mbox{\rm GeV}}^2)=2.776$~\cite{ICHEP} gives a
very small value of the scale parameter $\Lambda^{(n_f=4)}_{\rm PT}=89$~MeV.
The choice of normalization point influences the value of the parameter
$\Lambda$, but does not change the general picture of there being a
difference between the APT and PT results for low $Q^2$ scales.
To study this difference we take the relevant value of
$\Lambda^{(4)}=300$~MeV.
That practically means the normalization on the
value $S_{\rm GLS}(Q^2=3.16\,{\mbox{\rm GeV}}^2)=2.547$~\cite{ICHEP}
that gives $\Lambda_{\rm PT}^{(4)}=303$~MeV.
This value of the scale parameter $\Lambda$ is quite realistic and agrees
well with the results of the fits to the structure function $xF_3$ from
the CCFR'97 data, for example, with the two-loop result
$337\pm28$~MeV~\cite{Seligman97} and with the three-loop value
$308\pm34$ MeV~\cite{Kataev98}.
Since the difference between the PT and APT forms of the QCD corrections
is of order $\Lambda^2/Q^2$, both these functions will coincide in the
asymptotic region, where the perturbative approximation is valid.

\begin{figure}[thp]
~\hspace*{1.5cm}
\centerline{ \epsfig{file=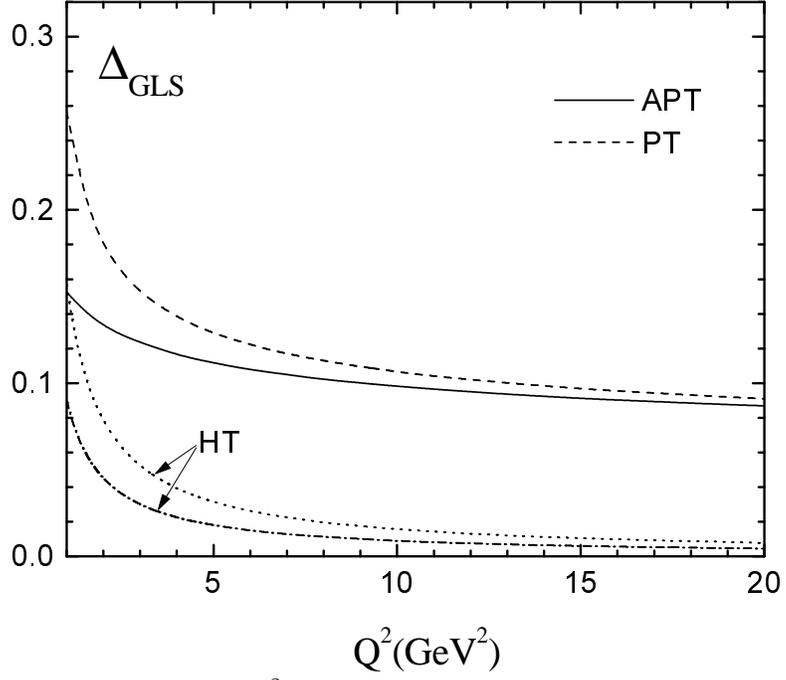,width=15.0cm}}
\caption{The theoretical predictions for $Q^2$
evolution of QCD correction to the GLS sum rule given by
Eqs.~(\protect\ref{Delta_PT}) and (\protect\ref{Delta_APT}).
The dash-dotted and dotted lines represent the HT corrections from
Refs.~\protect\cite{BBK,Ross}, respectively. }
\label{gls_fig3}
\end{figure}

The comparison between the $Q^2$ evolution in the APT and PT approaches
is shown in Fig.~\ref{gls_fig3}, where
the QCD correction to the GLS sum rule is plotted for
the perturbative part (solid curve for APT and dashed for PT approach) and
separately for the HT term given by two different
estimations for the coefficients
in the form  $(0.09\pm0.045)/Q^2$ (dash-dotted line)
and $(0.16\pm0.01)/Q^2$ (dotted line) taken from Refs.~\cite{BBK,Ross},
respectively. This figure demonstrates that there is an
essential difference between PT and APT evolutions for low $Q^2$.
Instead of a rapidly changing function with unphysical singularities as
occurs in the PT case, we get a slowly changing function in the APT approach.

\begin{figure}[thp]
~\hspace*{1.0cm}
\centerline{ \epsfig{file=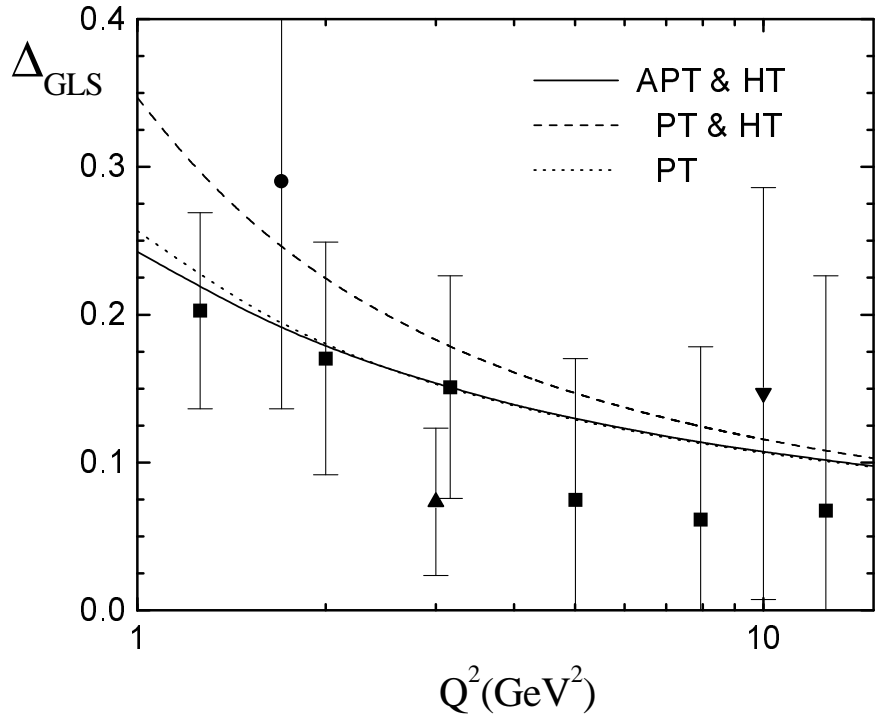,width=16.0cm}}
\caption{
The APT and PT predictions for the QCD correction to the GLS sum rule
together with experimental results. The Serpukov data~\protect\cite{Barabash}
is denoted by a circle, the CHARM~\protect\cite{CHARM} data by a downward
pointing triangle, the CCFR data from Ref.~\protect\cite{Oltman} by an upward
pointing triangle, and the recent CCFR/NuTeV results from
Ref.~\protect\cite{ICHEP} by squares.
}
\label{gls_fig4}
\end{figure}

In Fig.~\ref{gls_fig4} we plot the full contribution to the
QCD correction $\Delta_{\rm GLS}$ with the perturbative part,
calculated in the PT and APT approaches, and the HT
part taken in the form as discussed above~\cite{PDG96}.
We also plot the data, indicated by squares,
of the recent CCFR/NuTeV analysis~\cite{Yu2,ICHEP}
where the $Q^2$ behavior of the GLS integral has been
evaluated at values of $Q^2$ between $1.26\,{\mbox{\rm GeV}}^2$
and $12.59\,{\mbox{\rm GeV}}^2$, as well as older data.
The dotted curve represents the PT result without HT effects.
At high $Q^2$ scales, the PT and APT results agree closely with each other,
both including the HT effects and without them,
whereas at low $Q^2$ scales, the difference between the PT and APT
behaviors becomes significant.
At the same time, for $Q^2 > 2\,{\mbox{\rm GeV}}^2$, the PT
prediction without the HT term practically coincides with the APT curve
including the HT term.
Cancellation between the additional APT terms
beyond the standard PT prediction and the HT terms can explain the fact
that attempts to extract the HT effect from the CCFR'97 $xF_3$ experimental
data give a rather small value, even poorly determining the sign of the HT
contribution (see, for detail, Ref.~\cite{Kataev98}).
Note also that in the analysis of Refs.~\cite{Spentzouris,Yu1} the
HT contribution to the QCD correction for the GLS sum rule is taken to be
given by the form $(0.09\pm0.05)/Q^2$; however, in the subsequent
papers~\cite{Yu2,ICHEP,Harris_2},
the HT term has been taken to be smaller, $(0.05\pm0.05)/Q^2$.
Thus, the comparison of the APT and PT predictions with experimental data,
as is demonstrated in Fig.~\ref{gls_fig4}, cannot give a definite
conclusion since there are large experimental errors and
the value of the HT corrections is very uncertain.

We have considered the $Q^2$ evolution of the GLS sum rule in the
customary $\overline{\rm MS}$ renormalization scheme.
In general, it is not sufficient to obtain a result
in some scheme, but rather it is important to study its
RS stability over some acceptable domain.
The RS dependence of the GLS sum rule
based on the PT approach has been studied in Ref.~\cite{CK92}.
In the next section we consider the RS stability of the APT results.

\section{Renormalization scheme dependence}

A truncation of a perturbative expansion leads to
uncertainties in the theoretical predictions  arising
from the RS dependence of the partial sum of the series.
At low momentum scales these uncertainties may become very large
(see, for example, an analysis in Ref.~\cite{Raczka96}).
A physical quantity, in our case the QCD correction to the GLS sum rule,
has to be invariant under a change of RS, when the coupling constant
transforms as follows ($a=\alpha_S/\pi$)
\begin{equation}
\label{aRS}
a'=a\,(1+v_1a+v_2a^2+\cdots)\,.
\end{equation}
In the new RS, the QCD correction $\Delta_{\rm GLS}$ is represented as
\begin{equation}
\label{Delta-new}
\Delta_{\rm GLS}=a'\,(1+d_1'a'+d_2'{a'}^2)\, ,
\end{equation}
where the coefficients $d_1$ and $d_2$ are RS dependent
and have the transformation law
\begin{eqnarray}
\label{d_1-d_2}
d_1'&=&d_1-v_1\, , \\ \nonumber
d_2'&=&d_2-2(d_1-v_1)v_1-v_2\, .
\end{eqnarray}
In the three-loop level there are two process-dependent invariants
under the RS-\-trans\-for\-mation~(\ref{aRS})~\cite{Stevenson81}
\begin{eqnarray}
\label{RS-invar1}
\rho_1&=&\frac{b}{2}\ln\frac{Q^2}{\Lambda^2}-d_1 \, , \\
\rho_2&=&b_2+d_2-b_1d_1-d_1^2 \, ,
\label{RS-invar2}
\end{eqnarray}
where $b=\beta_0/2$, $b_1$ and $b_2$ are the coefficients of
the three-loop $\beta$-function
\begin{equation}
\label{beta}
\beta(a)=\mu^2\frac{\partial\,a}{\partial\,\mu^2}=-\beta_0\,a^2\,
(1+b_1a+b_2a^2)\, .
\end{equation}
The coefficient $b_2$, as well $d_1$ and $d_2$, depends on RS, and,
because of the first equation in (\ref{d_1-d_2}),
the scale parameter $\Lambda$ transforms as follows~\cite{Celmaster}:
$\Lambda'=\Lambda\,\exp(v_1/b)$.
In terms of the scale parameter $\Lambda_{\overline{\rm MS}}$, the
three-loop perturbative running coupling in any RS obeys the equation
\begin{equation}
\label{eq-to-a}
\frac{b}{2}\ln\left(\frac{Q^2}{\Lambda_{\overline{\rm MS}}^2}\right)=
d_1^{\overline{\rm MS}}-d_1+\Phi(a,b_2)\, ,
\end{equation}
where
\begin{eqnarray}
\label{Phi}
\Phi(a,b_2)&=&\frac{1}{a}-b_1\ln\frac{1+b_1a}{\beta_0\,a} \\ \nonumber
&+&b_2\int_0^a\frac{dx}{(1+b_1x)(1+b_1x+b_2x^2)}\, .
\end{eqnarray}

Thus, any RS taken from the ${\rm MS}$-like schemes can be characterized
by two parameters, which we choose here to be $d_1$ and $b_2$.
To calculate the second RS invariant one can use the coefficients
in the $\overline{\rm MS}$ scheme, which for
$n_f=4$ are $d_1=3.25,\, b_2=3.05\,$.

In the framework of the conventional approach
there is no solution of the RS dependence problem apart from
calculating many further terms in the asymptotic PT expansion, and there
is no fundamental principle upon which one can choose one
or another preferable RS. Usually, one uses
a class of `natural' or `well-behaved' RSs, which are defined
by the so-called cancellation index criterion~\cite{Raczka95},
according to which the degree of cancellation between the different terms
in the second RS-invariant $\rho_2$ are not too large,
as measured by the cancellation index
\begin{equation}
\label{CI}
C=\frac{1}{|\rho_2|}\left(  |b_2|+|d_2|+d_1^2+|d_1|\,b_1  \right)\, .
\end{equation}

By taking some maximum value of the cancellation index $C_{\rm max}$
one can investigate the stability of predictions for those RSs
with $C\leq C_{\rm max}\,$. In the case of the $\overline{\rm MS}$-scheme
the value of the cancellation index for the GLS sum rule
is $C_{\overline{\rm MS}} =95\,$.
Taking into account that the $\overline{\rm MS}$ scheme is commonly used,
we will consider this value as a boundary for the class of `natural' schemes.
In addition we will compare our results with predictions given by
optimized schemes based on the principle of minimal sensitivity
(PMS)~\cite{Stevenson81}
and the method of effective charge (ECH)~\cite{Grunberg84}.
(See also applications in~\cite{CK92,Mattingly,KataevRS}).

It should be stressed that the large value of the cancellation index
$C_{\overline{\rm MS}}\,$ for four active quark flavors
does not mean that the coefficients from which the
second RS invariant is constructed are huge, only that
$\rho_2$ is small, $\rho_2=-0.32\,$. In this special case the value of
$\rho_2$ is very sensitive to the value of $n_f$; for example, for
$n_f=3$, this invariant becomes $\, \rho_2^{(n_f=3)}=4.2\,$
and the cancellation index becomes significantly reduced.
At the same time, the absolute values of the coefficients that construct
the second RS invariant~(\ref{RS-invar2}) are smaller for $n_f=4$ than for
$n_f=3$, however, $C_{\overline{\rm MS}}^{(n_f=4)} \simeq 10\,
C_{\overline{\rm MS}}^{(n_f=3)}$.
Perhaps, in this situation, it is more convenient to introduce as the
corresponding index, which will define the class of `natural' RSs,
the sum of the absolute values of the terms in $\rho_2$,
without the denominator in Eq.~(\ref{CI}). The numerical parameter
introduced in such a way will
practically define the same region in the $(d_1,b_2)$-plane,
without changing significantly with changing $n_f$, and,
therefore, more adequately describing the situation.

Let us briefly discuss results of the ECH and PMS approaches.
For the ECH scheme the parameters are
$\,d^{\rm ECH}_1=d^{\rm ECH}_2\equiv 0$ and, therefore,
$b^{\rm ECH}_2 =\rho_2$. The transformation from the
$\overline{\rm MS}$ scheme  to the ECH scheme is performed with
the parameters in Eq.~(\ref{aRS})
$v_1^{\rm ECH}=d_1^{\overline{\rm MS}}=3.25$ and
$v_2^{\rm ECH}=d_2^{\overline{\rm MS}}=12.20$.
The system of equations for getting the PMS optimal prescription
consists of four equations.
These are Eqs.~(\ref {RS-invar2})  and (\ref{eq-to-a})
complemented by the following two equations:
\begin{eqnarray}
&&
3d_2+2d_1b_1+b_2+(2d_1b_2+3d_2b_1)a+3d_2b_2a^2=0\, ,     \nonumber \\
&&
(1+b_1a+b_2a^2)(1+2d_1a+3d_2a^2)I(a,b_2)-a=0\, .
\end{eqnarray}
where
\begin{equation}
\label{Iab_2}
I(a,b_2)=\int_0^a\frac{dx}{(1+b_1x+b_2x^2)^2}\, .
\label{PMS}
\end{equation}
In the PMS optimization procedure
the coefficients $d_1$, $d_2$ and $b_2$ become $Q^2$ dependent and,
therefore, have to be adjusted for each different value of $Q^2$.

\begin{figure}[thp]
~\hspace*{1.0cm}
\centerline{ \epsfig{file=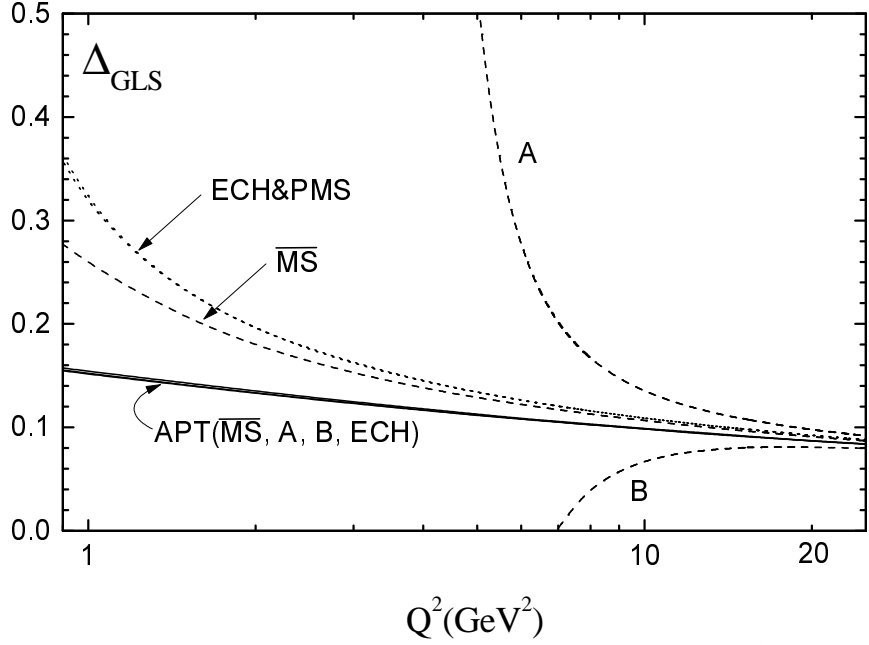,width=14.0cm}}
\caption{Renormalization scheme dependence of predictions for
$\Delta_{\rm GLS}$ vs. $Q^2$ for the APT and PT expansions.
The solid curves, which are very close to each other,
correspond to the APT results for the $\overline{MS}$, A, B,
and ECH schemes.
The PT predictions in the $\overline{MS}$,~A, and B schemes are
represented by the dashed curves;
the ECH and PMS results are given by the dotted lines.}
\label{gls_fig5}
\end{figure}

\begin{figure}[thp]
~\hspace*{-1.0cm}
\centerline{ \epsfig{file=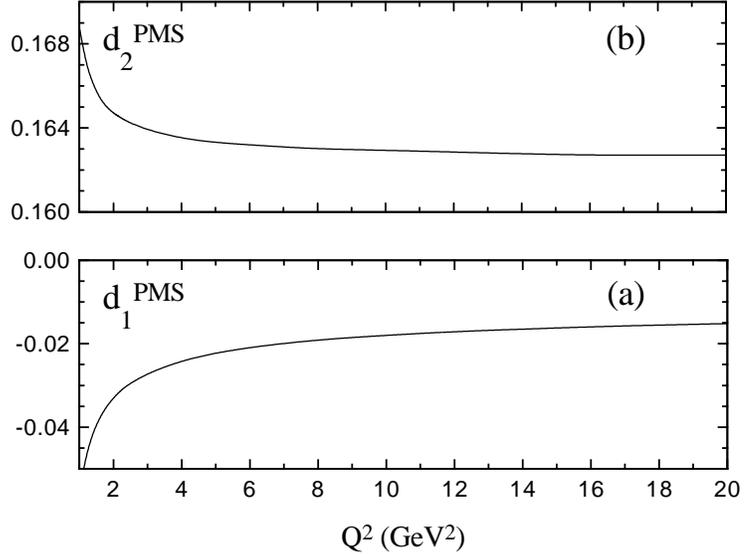,width=8.0cm}}
\caption{The $Q^2$ dependence of the PMS coefficients
for the GLS sum rule, $n_f=4$.}
\label{gls_fig6}
\end{figure}

We present our results in Fig.~\ref{gls_fig5},
where the QCD correction  $\Delta_{\rm GLS}$ is plotted
as a function of $Q^2$ in different RS. The schemes A and B have the
same cancellation index as in the $\overline{\rm MS}$ scheme, $C=95$,
and are defined by the following values of the parameters:
$d_1^{\rm A}=-3.93$, $b_2^{\rm A}=0$ and
$d_1^{\rm B}=-1.0$, $b_2^{\rm B}=13.7$.
The dotted curves present the ECH and PMS predictions, which
are very close to each other. This is because the PMS coefficients
$d_1^{\rm PMS}(Q^2)$ and $d_2^{\rm PMS}(Q^2)$
are found for four active quark flavors
to be small numerically in the region under consideration,
$Q^2 > 1\,{\rm GeV}^2$.
Behaviors of these coefficients as functions of $Q^2$
are shown separately in Fig.~\ref{gls_fig6}.

\begin{figure}[thp]
\centerline{ \epsfig{file=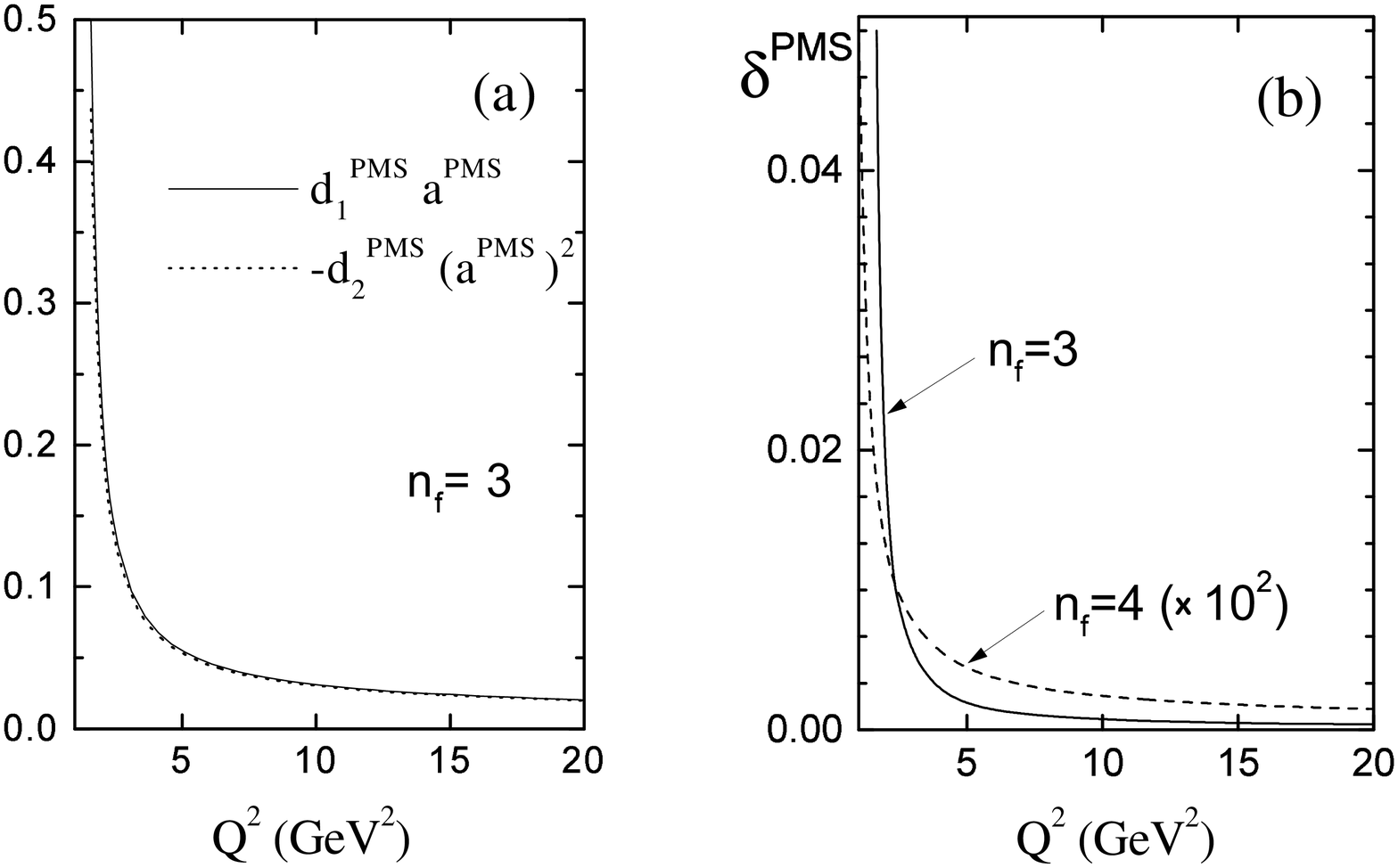,width=12.0cm}}
\caption{Cancellation in the PMS optimal prescription vs.~$Q^2$:
a)~quadratic and cubic contributions
for $\Delta_{\rm GLS}=a^{\rm PMS}
\left[ d_1^{\rm PMS}a^{\rm PMS}+
d_2^{\rm PMS}\left(a^{\rm PMS}\right)^2 \right]$,  $n_f=3$;
b)~$\delta^{\rm PMS} =d_1^{\rm PMS}a^{\rm PMS} +
d_2^{\rm PMS}\left(a^{\rm PMS}\right)^2\,$.
}
\label{gls_fig7}
\end{figure}

As it has been mentioned above, there are no strong arguments to fix
the definite number of active quark flavors for the integral~(\ref{GLS})
to be $n_f=4$. Therefore, we have
also considered the case  $n_f=3$. The ECH and PMS prescriptions
gave again results that are very close to each other.
However, the reason is no longer that the PMS coefficients
$d_1^{\rm PMS}(Q^2)$ and $d_2^{\rm PMS}(Q^2)$
are very small, but that there is a strong cancellation
between the last two terms in the
expression for the QCD correction
$\Delta_{\rm GLS}=a^{\rm PMS}(1+\delta^{\rm PMS})$ with
$\delta^{\rm PMS}\equiv d_1^{\rm PMS}a^{\rm PMS}+
d_2^{\rm PMS}\left(a^{\rm PMS}\right)^2\,$.
This fact is demonstrated in Fig.~\ref{gls_fig7}, where
(a) shows the quadratic (solid line) and cubic
(dotted line) contributions to $\Delta_{\rm GLS}$ for $n_f=3$.
The sum of these contributions is shown in Fig.~\ref{gls_fig7}~(b)
for  $n_f=3$ (solid line) and for $n_f=4$ (dashed line).

The APT predictions for the $\overline{\rm MS}$, A, and B schemes
turn out to be stable for the whole interval of momenta and appears as a
wide solid line in Fig.~\ref{gls_fig5}. It should be stressed that the
optimized RS constructions in the framework of PT do not maintain
the correct causal properties, however, to avoid this difficulty,
they can be modified in the sense of APT. In Fig.~\ref{gls_fig5}
we plot the analytic ECH result, which, of course, is different from the
perturbative one and practically coincides with the other APT curves
for the $\overline{\rm MS}$, A, and B schemes.

\begin{figure}[thp]
~\hspace*{1.5cm}
\centerline{ \epsfig{file=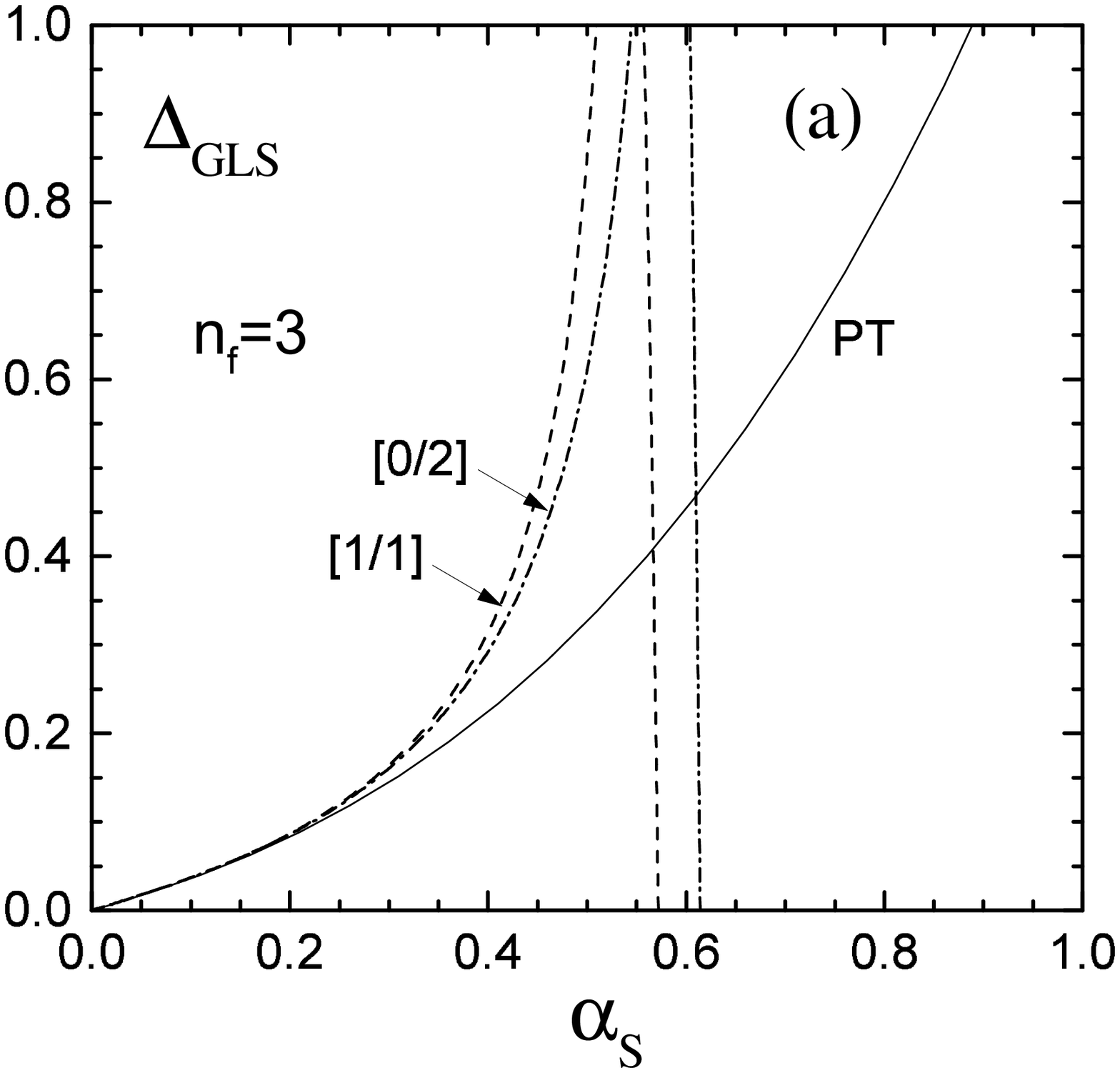,width=12.0cm}
~\hspace*{-4.0cm}
\epsfig{file=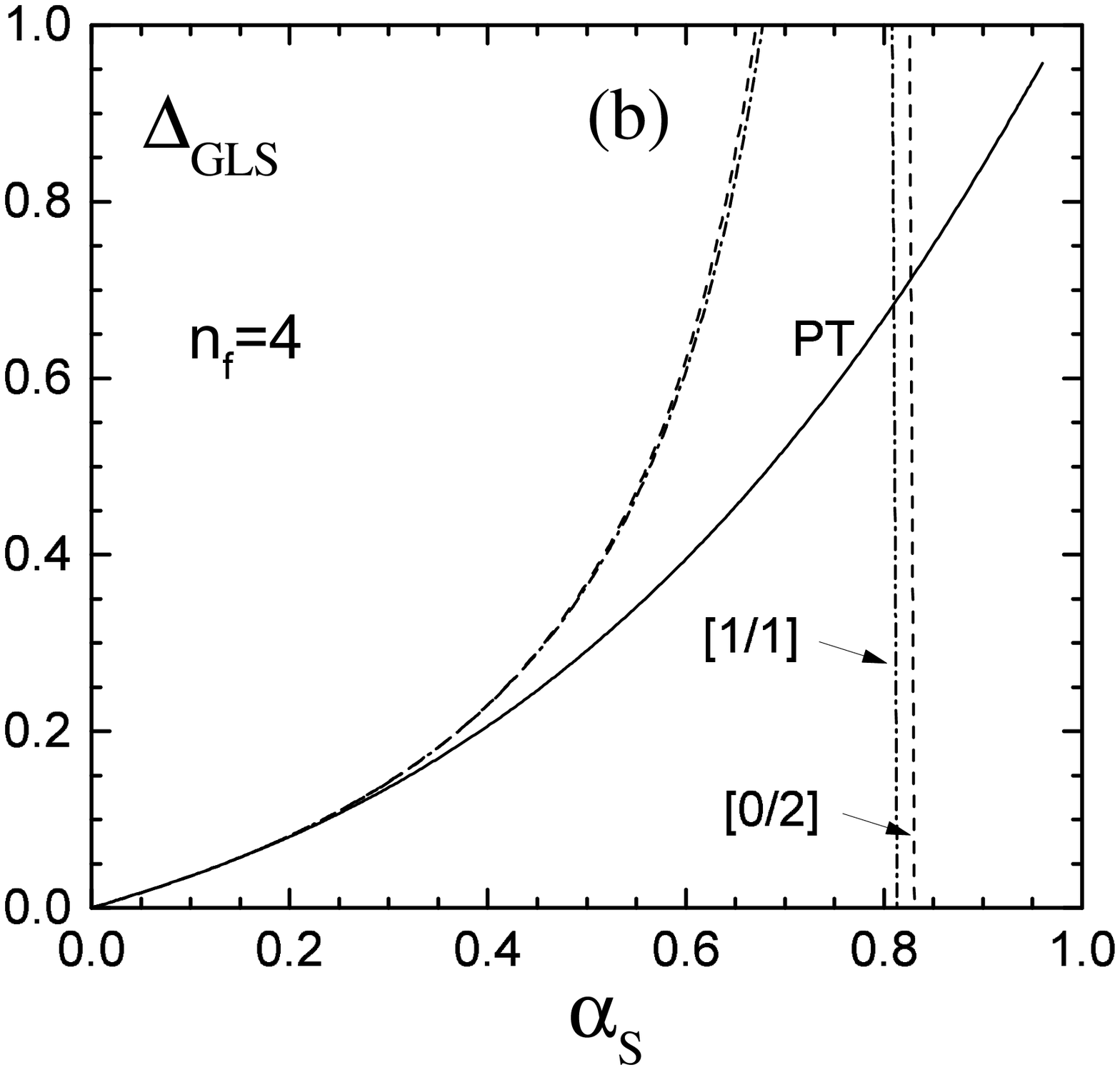,width=12.0cm}}
\caption{Behavior of the PA [1/1] and [0/2]
for QCD corrections to GLS sum rule vs.~$\alpha_{\rm S}$:
a) $n_f=3$ ; b) $n_f=4\,$ .
}
\label{gls_fig8}
\end{figure}

The sensitivity of the PT predictions to the RS dependence can be reduced
by applying the Pad$\acute{\rm e}$ approximation (PA)~\cite{Pade}. However,
in the case under consideration, this approach can lead to some difficulties.
In Fig.~\ref{gls_fig8} we plot two possible
Pad$\acute{\rm e}$ approximants $[1/1]$ and $[0/2]$ in the three-loop order
for $\Delta_{\rm GLS}$
versus $\alpha_S$. This figure demonstrates that the PA leads to
unphysical singularities in the region of small momenta that violate
the causal properties and prevents us from considering the method of PA as a
systematic approach.  In contrast, the APT results
for the $\overline{\rm MS}$, A, B, and ECH schemes lead to stable predictions
for the whole interval of momentum, which practically coincide with each
other, and, therefore, the APT description turns out to be practically
RS independent.

\section{Summary and conclusion}

We have considered the GLS sum rule by using the APT
approach, which maintains the causal properties and, in contrast to PT,
does not lead to any unphysical singularities.
Taking into account the analytic properties of the moments of the deep
inelastic structure functions, we have obtained an
analytically improved theoretical description of the GLS sum rule.
We have shown that the convergence properties of the APT expansion
are better than are those in the standard PT description.
We have demonstrated further that there is a significant difference between
the $Q^2$ evolution in the PT and APT approaches for low $Q^2$ scales.
We have also analyzed the GLS sum rule taking into account
the higher-twist contribution. The APT power corrections have a sign
opposite to that of the typically-used higher-twist term~\cite{PDG96}
and, numerically, there is an effective cancellation
between these two corrections.

In this paper we have found that in the framework of APT the theoretical
prediction to the GLS sum rule becomes practically renormalization
scheme independent starting from the three-loop level.
A similar statement has been made for the $e^+e^-$ annihilation
into hadrons~\cite{SDVRS} and for the semileptonic inclusive decay
of the $\tau$ lepton~\cite{MSYa}. A corresponding analysis of the Bjorken
sum rule is presented in Ref.~\cite{MSS2}. Thus, the APT approach gives a
systematic method of reducing the RS ambiguity significantly and leads to
practically unique predictions for physical quantities.
It should be stressed that
our analysis shows that there is serious doubt concerning the conjecture
that, in spite of the large values of $\alpha_{S}$ at low $Q^2$,
the conventional three-loop PT predictions for the GLS sum rule are reliable
(see, e.g.,~\cite{Harris_2}).
The proximity of PT predictions in the $\overline{\rm MS}$ scheme
to predictions obtained in the optimal schemes seems to possess no significance
for any process at small momentum transfer and cannot be considered
a guarantee of RS independence of the results.

Unfortunately, at present, the experimental situation
for the GLS sum rule, with its large errors, does not allow us
to come to a reliable conclusion that the
APT description is preferable to the PT one. However, from the theoretical
point of view, the remarkable properties of the APT approach,
for example, the maintenance of causality and the higher-loop and
renormalization-scheme stability for the whole interval of momenta,
create a basis for preferring the application of this new technique.

\section*{Acknowledgement}

The authors would like to thank L.~Gamberg, A.~L.~Kataev,
and S.~M.~Mikhailov for useful discussions and interest in this work,
O.~Nachtmann who brought the paper~\cite{W78} to our attention,
and J.~H.~Kim for useful discussion of the  CCFR/NuTeV data.

Partial support of the work by the US National Science Foundation,
grant PHY-9600421,  by the US Department of Energy,
grant DE-FG-03-98ER41066, by the University of Oklahoma,
and by the RFBR, grant 96-02-16126 is gratefully acknowledged.


\end{document}